\documentclass[aps,prb,reprint,showpacs,floatfix]{revtex4-1}
\usepackage{graphicx}  
\usepackage{dcolumn}   
\usepackage{bm}        
\usepackage{amssymb}   
\usepackage{amsmath, amsthm}
\usepackage[extra]{tipa}
\usepackage{color}
\setcitestyle{square,numbers}

\begin{document}
\title{Luttinger liquid and charge-density-wave phases in
a spinless fermion wire on a semiconducting substrate}
\author{Anas Abdelwahab}
\author{Eric Jeckelmann}
\affiliation{Leibniz Universit\"{a}t Hannover, Institut f\"{u}r Theoretische Physik, Appelstr.~2, 30167 Hannover, Germany}

\date{\today}

\begin{abstract}
An interacting spinless fermion wire coupled to a three-dimensional (3D) semiconducting substrate is approximated by a
narrow ladder model (NLM) with varying number of legs. We compute density distributions, gaps, charge-density-wave (CDW) order parameters,
correlation functions, and the central charge using the density-matrix renormalization group method.
Three ground-state phases are observed: a one-component
Luttinger liquid, a quasi-one-dimensional (1D) CDW insulator, and a band insulator.
We investigated the convergence of the NLM properties with increasing number of legs systematically
and confirm that the NLM is a good approximation for the quasi-1D phases (Luttinger liquid and CDW) of the 3D wire-substrate model.
The quantum phase transitions between these phases are investigated as function of the coupling between wire and substrate.
The critical nearest-neighbor interaction increases with increasing coupling between wire and substrate
and thus the substrate stabilizes the Luttinger liquid in the wire.
Our study confirms that a Luttinger liquid or CDW insulator phase could occur in the low-energy properties of atomic wires deposited on
semiconducting substrates. 
\end{abstract}

\maketitle

\section{\label{sec:intro}Introduction}

Atomic wires on semiconducting surfaces are prime candidates to realize the physics
of one-dimensional (1D) correlated electrons~\cite{springborg07,onc08,sni10} but the interpretations of 
experimental results are highly controversial. For instance, Au/Ge(001)~\cite{blu11,blum12},
Bi/InSb(100)~\cite{ohts15} and Pt/Ge(100)~\cite{yaji13,yaji16} have been described as quasi-1D conductors
and thus as possible realizations of Luttinger liquids~\cite{Schoenhammer,giamarchi07,solyom}.
Charge-density-wave (CDW) states~\cite{gruener,chen} have been reported in In/Si(111)~\cite{yeom99,cheo15} and 
Au/Si(553)~\cite{shin,aulbach}. 
However, the theory of correlated electrons in quasi-1D systems is well established only for isolated
chains and narrow ladders, as well as for anisotropic bulk electronic systems~\cite{sol79,giamarchi07,gruener,solyom}.
It has not been extended yet to account for
the influence of a three-dimensional (3D) host such as a semiconducting substrate on the properties of
a 1D Luttinger liquid or an electronic CDW. 

The physical properties of Luttinger liquids and CDW systems can be studied theoretically
using lattice models.
In principle, the properties of quasi-1D lattice models can be calculated using
the density matrix renormalization group (DMRG) method~\cite{whi92,whi93,sch05,jec08a}. 
However, DMRG cannot treat 3D lattice models for wire-substrate systems directly.
Therefore, in a previous publication~\cite{paper1}, we introduced a 3D lattice model
for a correlated atomic wire deposited on a substrate and showed how to map it exactly onto
a two-dimensional (2D) ladder-like lattice that can be approximated by quasi-1D narrow ladder models (NLM)
with increasing number of legs.
We demonstrated the approach using the 1D Hubbard model to represent a correlated atomic wire~\cite{paper2}.
Due to the high computational cost of DMRG for electronic ladder systems, we were not able to
study the convergence (and thus the stability of 1D features) with the number of legs systematically. 

In this paper, we apply the NLM approach to a correlated wire represented by
the 1D spinless fermion (1DSF) model~\cite{giamarchi07,solyom}. 
This model is defined by the Hamiltonian
\begin{eqnarray}
\label{eq:1Dspinlessmodel}
 H &=& -t_{\text w}  \sum^{L_x-1}_{x} \left ( c^{\dag}_{x}  
c^{\phantom{\dag}}_{x+1} + \text{H.c.} \right ) \\ \nonumber
 && + V \sum^{L_x-1}_x \left ( c^{\dag}_{x+1}c^{\phantom{\dag}}_{x+1} - \frac{1}{2} \right)
 \left ( c^{\dag}_{x}c^{\phantom{\dag}}_{x} - \frac{1}{2} \right)
\end{eqnarray}
where $c_{x}$ ($c^{\dag}_{x}$) annihilates (creates) a spinless fermion on site $x$ residing
in a 1D lattice with length $L_x$. 
The parameter $t_{\text w}$ determines the hopping amplitude between nearest-neighbor sites while 
$V$ determines the strength of the Coulomb interaction between nearest-neighbor fermions.
As the model exhibits a particle-hole symmetry that changes the sign of the hopping term only, 
we can assume without loss of generality that $t_{\text w}\geq 0$.

This model is exactly solvable using the Bethe Ansatz method and its properties are well known~\cite{gaudin,giamarchi07}.
Here we focus on the repulsive case $V\geq 0$ and thus the (grand-canonical) ground state
occurs at half filling, i.e. for $N=L_x/2$ spinless fermions on the lattice.
The model exhibits two different ground-state phases at half-filling as a function of $V\geq 0$.  
In the range $V \leq 2t_{\text w}$ the ground-state density distribution is uniform,
$n_x=\left \langle c^{\dag}_{x}c^{\phantom{\dag}}_{x} \right \rangle=\frac{1}{2}$.
The excitation spectrum is gapless and its low-energy sector
is described by a one-component Luttinger liquid.  
For $V>2t_{\text w} $, the 1DSF model exhibits a spontaneous broken-symmetry ground state
with a CDW $n_x= \frac{1}{2} + (-1)^x  \delta$ ($0< \vert \delta \vert < \frac{1}{2}$) 
while its excitation spectrum is gapped. 
$V_{\text{CDW}}=2t_{\text w}$ is the quantum critical point of the 
continuous quantum phase transition
between the CDW and the Luttinger liquid phases.

In this paper we investigate the fate of theses phases when the wire is coupled to a semiconducting substrate.
DMRG allows us to compute broader ladder systems for spinless fermions than for electronic models. 
For this study we have used NLM with up to 15 legs. The slow increase of entanglement with the number of legs
in the NLM~\cite{paper2} allows us to study large ladder widths with high accuracy and reasonable computational cost.
Therefore, we can perform a more accurate study of the convergence with the number of legs and
confirm that the NLM approach can describe the quasi-1D low-energy physics occurring in 3D wire-substrate systems.
We demonstrate that Luttinger liquids and CDW states remain stable when coupled to a non-interacting gapped substrate
and thus shed some light on these hallmarks of 1D correlated electron systems 
in atomic wires deposited on semiconducting substrates.

\section{\label{sec:models}Models}

\subsection{3D wire-substrate model \label{sec:full_model}}

We start from a 3D wire-substrate model that is similar to the one introduced in our previous work~\cite{paper1}.
However, we consider only spinless fermions and the 1DSF Hamiltonian~(\ref{eq:1Dspinlessmodel})
is substituted for the 1D Hubbard Hamiltonian. 
The full Hamiltonian takes the form
\begin{eqnarray}
\label{eq:hamiltonian}
 H&=&
 -t_{\text w}  \sum_{x} \left ( c^{\dag}_{{\text w} x}  
c^{\phantom{\dag}}_{{\text w},x+1} + \text{H.c.} \right ) \nonumber \\
&& + V \sum_x \left ( c^{\dag}_{{\text w}x+1}c^{\phantom{\dag}}_{{\text w}x+1} - \frac{1}{2} \right)
 \left( c^{\dag}_{{\text w}x}c^{\phantom{\dag}}_{{\text w}x}  - \frac{1}{2} \right) \nonumber \\
&& + \sum_{b, \bm{r} }  \epsilon_{\text b}  c^{\dag}_{{\text b}\bm{r}}  c^{\phantom{\dag}}_{{\text b}\bm{r}}
-t_{\text s} \sum_{\langle \bm{r} \bm{r'} \rangle} \sum_{\text{b} } \left (
c^{\dag}_{{\text b}\bm{r} }  c^{\phantom{\dag}}_{{\text b}\bm{r'}} + \text{H.c.}
\right ) \nonumber \\
&& -t_{\text{ws}} \sum_{b, <x,r>} \left ( c^{\dag}_{{\text b} \bm{r} }  
c^{\phantom{\dag}}_{{\text w} x } + \text{H.c.}  \right ) .
\end{eqnarray}

The substrate lattice is a cubic lattice of size $L_x \times L_y\times L_z$ with
open boundary conditions in the $z$-direction and periodic boundary conditions in the $x$ and $y$ directions.
The sum over $\bm{r}$ runs over all substrate lattice sites
and the sum over $\langle \bm{r} \bm{r'} \rangle$ is over all pairs of nearest-neighbor sites in the substrate. 
The substrate is modeled by a tight-binding Hamiltonian with nearest-neighbor hopping $t_{\text s}$ 
and two orbitals per site, one for the valence band and one for the conduction band. 
The operator $c^{\dag}_{{\text b} \bm{r} }$ creates a spinless fermion on the site with coordinates $\bm{r} = (x,y,z)$
in the valence ($\text{b}=\text{v}$) or conduction ($\text{b}=\text{c}$) orbital.
In momentum space the single-particle dispersions take the form
\begin{equation}
\label{eq:disp}
\epsilon_{\text{b}}(\bm{k})  = \epsilon_{\text b} - 2t_{\text s} [ \cos(k_x) + \cos(k_y) + \cos(k_z) ] ,
\end{equation} 
where $k_x, k_y \in [-\pi,\pi]$ and $k_z \in [0,\pi]$ while  
$\epsilon_{\text b}=\pm \epsilon_{\text s}$ denotes the on-site energies for the valence
and conduction ($\text{b}=\text{v,c}$) bands. 
Thus there is an indirect gap 
$\Delta_{\text s} =  2 \epsilon_{\text s} - 12 t_{\text s} $
between the bottom of the conduction band and the top of the valence band.
The spectrum is gapped only if $\epsilon_{\text s} > 6 t_{\text s}$
and this condition must be fulfilled to represent a semiconducting substrate.

The wire is aligned with the substrate surface in the $x$-direction at the position $y=y_0 \in \{1,\dots,L_y\}$ and $z=0$.
The operator $c^{\dag}_{{\text w} x}$ creates a spinless fermion on the wire site at the position $\bm{r} = (x,y_0,0)$. 
The last term in~(\ref{eq:hamiltonian}) represents the hybridization
between the wire and the substrate which is a single-particle hopping $t_{\text{ws}}$ between each wire site and the adjacent
substrate site at the position $\bm{r} = (x,y_0,1)$.
The sums over $x$ run over all wire sites from $1$ to $L_x$. 

Note that the Hamiltonian~(\ref{eq:hamiltonian}) describes a single wire.
Real systems of atomic wires on semiconducting substrates are made of several parallel
wires.
Thus we assume here that the (direct or substrate-mediated) interactions between wires
can be neglected. This is justified in first approximation for Luttinger liquids and ground-state CDW phases.
Several parallel wires would have to be taken into account to study quasi-1D long-range ordered phases 
at finite temperature, however.

\subsection{Narrow ladder model \label{sec:nlm}}

Applying the exact mapping introduced in our previous work~\cite{paper1} to the Hamiltonian~(\ref{eq:hamiltonian}),
we get  a ladder-like Hamiltonian on a 2D lattice of size $L_x \times M$ where $M=2L_yL_z+1$ is the number of legs
\begin{eqnarray}
\label{eq:ladder-hamiltonian}
 H&=&
 -t_{\text w}  \sum_{x} \left ( g^{\dag}_{x,0}  
g^{\phantom{\dag}}_{x+1,0} + \text{H.c.} \right ) \nonumber  \\
  && + V \sum_x \left( g^{\dag}_{x+1,0}g^{\phantom{\dag}}_{x+1,0} - \frac{1}{2} \right)
  \left( g^{\dag}_{x,0}g^{\phantom{\dag}}_{x,0} - \frac{1}{2} \right) \nonumber \\
  &&-t_{\text s} \sum^{M-1}_{n=1}\sum_{x}
  \left ( g^{\dag}_{xn}g^{\phantom{\dag}}_{x+1,n} +\text{H.c.}\right) \nonumber \\
  &&-\sum^{M-2}_{n=0} t^{\text{rung}}_{n+1}
  \sum_{x}\left( g^{\dag}_{xn}g^{\phantom{\dag}}_{x,n+1}+\text{H.c.}\right). 
\end{eqnarray}
Here, $g^{\dag}_{xn}$ creates a fermion at position $x$ in the $n$-th leg ($n=0,\dots,M-1$).
The first leg ($n=0$) is identical with the wire, in particular $g^{\dag}_{x0} = c^{\dag}_{\text{w}x}$,
while legs $n=1,\dots,M-1$ correspond to successive substrate shells around the wire.  
The Hamiltonian~(\ref{eq:ladder-hamiltonian}) consists of the original 1DSF model on the leg representing the
correlated atomic wire, 
an intra-leg hopping $t_{\text s}$ in every leg representing the substrate, and a nearest-neighbor rung hopping
$t^{\text{rung}}_n$ between substrate legs $n-1$ and $n$.
The first two rung hoppings  $t^{\text{rung}}_{1}=\sqrt{2}t_{\text{ws}}$ and
$t^{\text{rung}}_{2}=\sqrt{3t^2_{\text s}+\epsilon_{\text s}^2}$ can be obtained algebraically.
For larger $n$, $t^{\text{rung}}_{n+1}$ can be computed easily using the Lanczos algorithm as described
in Sec.~III of Ref.~\cite{paper1}. The precise relation between the operators $c^{\dag}_{{\text b} \bm{r}}$
and $g^{\dag}_{xn}$ is also explained there.

Obviously, this 2D model can be approximated by restricting the number of legs that are taken into account.
This corresponds to substituting $N_{\text{leg}} (\leq M)$ for $M$ in the Hamiltonian~(\ref{eq:ladder-hamiltonian}).
This approximation will become better when $N_{\text{leg}}$ increases up to $M$.
In our previous works~\cite{paper1,paper2} we showed that one can already obtain useful information 
about the low-energy properties of the Hubbard wire (such as correlation functions along the wire, excitations localized in or around the wire) 
using narrow ladders with $N_{\text{leg}} << M$. 
Here we apply this approach to the spinless fermion model~(\ref{eq:hamiltonian}).
As for the Hubbard chain we have found that the number of legs $N_{\text{leg}}$ must be an odd number
to represent correctly the valence and conduction bands.

Throughout this work we will take the energies in the unit of $t_{\text{s}}$.
The wire parameter is fixed to $t_{\text{w}}=3t_{\text{s}}$ and the on-site energy   
is fixed to $\epsilon_{\text{s}}=7t_{\text{s}}$. The resulting indirect substrate band gap
is $\Delta_{\text s} =  2 \epsilon_{\text s} - 12 t_{\text s}=2t_{\text{s}}$ in the absence of any
interaction with the wire.
The 1D band of the noninteracting ($V=0$) and uncoupled ($t_{\text{ws}}=0$) wire is centered around the middle of the substrate gap.
We focus on the half-filled band case where the number of spinless fermions is $N=L_x M/2 = L_xL_yL_z+L_x/2$ for the
3D model~(\ref{eq:hamiltonian}) or $N=L_x N_{\text{leg}}/2$ for a NLM with $N_{\text{leg}}$ legs. Due to the particle-hole symmetry of the
Hamiltonian, the Fermi energy is always equal to zero and thus in the middle of the wire band.
Moreover, the average number of spinless fermions on the wire is always equal to $L_x/2$ although the
actual number fluctuates significantly as soon as the hybridization with the substrate is turned on ($t_{\text{ws}}\neq 0$).
As discussed in our previous works~\cite{paper1,paper2} the effective substrate band gaps $\Delta_{\text{s}}(N_{\text{leg}})$
of the NLM are larger than the true substrate band gap $\Delta_{\text s}$ but converge to $\Delta_{\text s}$ for 
$N_{\text{leg}} \rightarrow \infty$. This overestimation of the substrate band gap is the leading 
cause for quantitative differences between the 3D wire-substrate model and its NLM approximations for $N_{\text{leg}} < M$.

\section{Methods \label{sec:dmrg}}

DMRG is a well-established numerical method for studying quasi-1D correlated quantum lattice systems
with short-range interactions~\cite{whi92,whi93,sch05,jec08a}.
For 2D systems and ladders, however, DMRG is limited by an exponential increase  
of CPU time and required memory as a function of the lattice width or the number of legs, respectively.
As discussed in Ref.~\cite{paper2}, this problem is less severe when applying DMRG to 
the NLM because the number of gapless excitation modes does not increase with the number of legs
when the substrate is gapped.   
This is reflected in the relatively slow growth of entanglement with the ladder width. 
Moreover, the smaller dimension of the site Hilbert space in a spinless fermion model reduces the required computational 
cost in comparison to the Hubbard model studied in Ref.~\cite{paper2}.

In the present work
the ground-state properties of NLM with up to $N_{\text{leg}}= 15$ legs were computed.
Obviously the ladder Hamiltonian~(\ref{eq:ladder-hamiltonian}) is not periodic in the rung direction.
We also used open boundary conditions in the leg direction ($x$-direction) 
and always took an even number of rungs up to $L_x=200$ for broad ladders and up to $L_x=500$ for three-leg ladders.
We used the finite-system DMRG algorithm with up to $m=512$ density-matrix eigenstates
kept in the calculations yielding discarded weights smaller than $10^{-6}$.
The truncation errors were investigated by varying the number of density-matrix eigenstates
and extrapolating ground-state energies to the limit of vanishing discarded weights~\cite{bonc00}.

\section{Results \label{sec:results}}

For $V=0$ the low-energy excitations of the half-filled NLM~(\ref{eq:ladder-hamiltonian})
are gapless and localized in or around the wire for any hybridization strength $t_{\text{ws}}$,
resulting in a 1D metal.
As mentioned in Sec.~\ref{sec:intro}, the ground state of the isolated half-filled 1DSF model (corresponding to $t_{\text{ws}}=0$)  
undergoes a continuous quantum phase transition
at $V_{\text{CDW}}=2t_{\text w}=6t_{\text s}$ from a Luttinger liquid to a gapped CDW for increasing interaction
$V$.
We will investigate the stability of these quasi-1D phases and their phase transition when the wire is hybridized with the substrate
($t_{\text{ws}}\neq 0$).  
First, we discuss the behavior of the single-particle gap $E_{\text{p}}$ 
as a function of the interaction strength $V$ and the wire-substrate hybridization $t_{\text{ws}}$
for finite-size NLM to give a general overview of the physics in this model. 
Finite-size effects, order parameter and correlation functions will be addressed in the subsequent subsections.

\begin{figure}[t]
\includegraphics[width=0.48\textwidth]{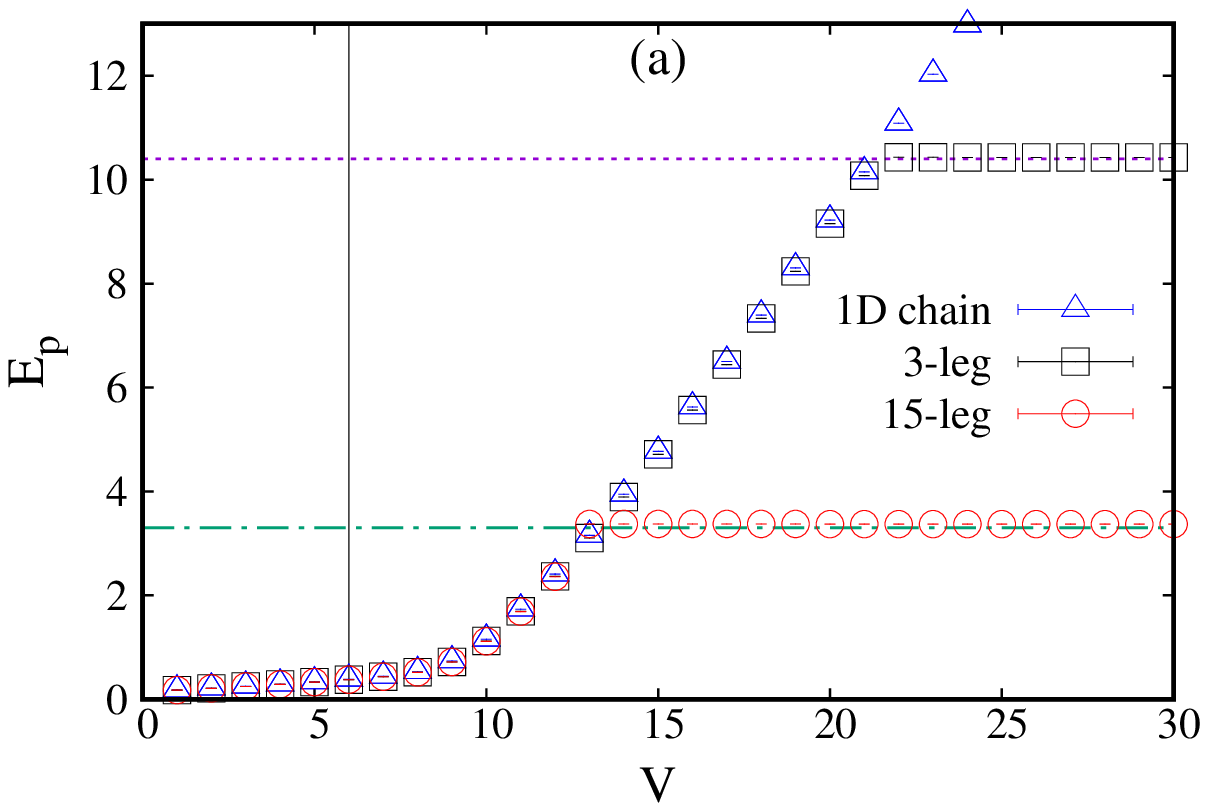}
\includegraphics[width=0.48\textwidth]{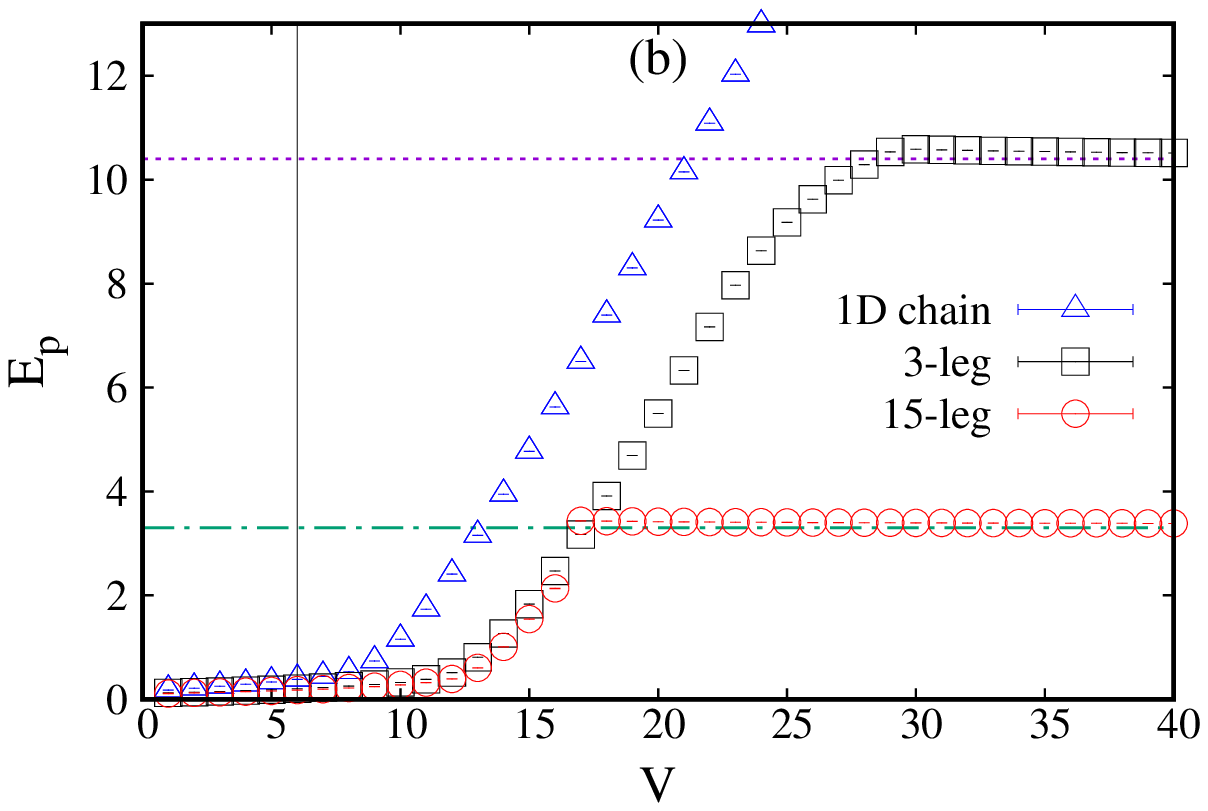}
\caption{\label{fig:Ep_vs_V_Ly}
Single-particle gap $E_{\text{p}}$ as a function of the nearest-neighbor interaction $V$
 for a 3-leg and a 15-leg NLM 
 as well as for the 1DSF model.
In all three cases the system length is $L_x=128$.
Panel (a) shows results for $t_{\text{ws}}=0.5t_{\text{s}}$ and
panel (b) for $t_{\text{ws}}=4t_{\text{s}}$.
Vertical lines show the critical coupling 
 $V_{\text{CDW}}=6t_{\text{s}}$ of the isolated spinless fermion chain in the thermodynamic limit
while horizontal lines shows the effective substrate band gaps $\Delta_{\text{s}}(N_{\text{leg}})$.
}
\end{figure}

The single-particle gap $E_{\text{p}}$ at half filling is defined as
\begin{equation}
 \label{eq:gap}
E_{\text{p}} = E_{0}(N+1)+E_{0}(N-1) - 2E_{0}(N)
\end{equation}
where $E_{0}(N')$ represents the ground-state energy of a ladder with $N'$ fermions and $N=L_xN_{\text{leg}}/2$.
This gap is shown in Fig.~\ref{fig:Ep_vs_V_Ly} for a 3-leg and a 15-leg NLM with $L_x=128$ for two values of the wire-substrate hybridization
$t_{\text{ws}}=0.5t_{\text{s}}$ and $4t_{\text{s}}$. 
We see that the gap $E_{\text{p}}$ increases monotonically with the interaction strength $V$
up to a value $V_{\text{BI}}$ where it abruptly saturates to a constant value, which is equal to
the effective band gap $\Delta_{\text{s}}(N_{\text{leg}})$ in the NLM, 
$\Delta_{\text{s}}(N_{\text{leg}}=3)\approx 10.4t_{\text{s}}$
and $\Delta_{\text{s}}(N_{\text{leg}}=15)\approx 3.3t_{\text{s}}$, respectively.

This saturation effect is similar to the one observed previously in the Hubbard wire-substrate model~\cite{paper1,paper2}
and has a similar origin. The lowest excitations are localized in the wire
as along as they are gapless or have a gap smaller than $\Delta_{\text{s}}(N_{\text{leg}})$.
However, the CDW gap increases with the interaction $V$ up to $\infty$ and thus always reaches the value $\Delta_{\text{s}}(N_{\text{leg}})$
for a finite coupling $V=V_{\text{BI}}$. For larger $V$ the low-energy excitations are delocalized in the substrate legs
and correspond to excitations from the valence band to the conduction band in the 3D wire-substrate model~(\ref{eq:hamiltonian}).
This interpretation is confirmed by the excitation density discussed in Sec~\ref{sec:density}.
Therefore, this phase is best described as a band insulator.

In Fig.~\ref{fig:Ep_vs_V_Ly}(a) we display $E_{\text{p}}$ as a function of $V$ for a
weak hybridization $t_{\text{ws}}=0.5t_{\text{s}}$. 
This gap is very close to the one obtained for the isolated 1DSF chain~(\ref{eq:1Dspinlessmodel}) 
up to $V_{\text{BI}}$.
Note that the finite gaps seen for weak $V$ are
due to finite-size effects and vanish in the thermodynamic limit
($1/L_x\rightarrow 0$) as shown in Sec.~\ref{sec:gap}. 
For intermediate coupling ($6t_{\text{s}} \alt V < V_{\text{BI}}$) the NLM exhibits a gap similar to the
one found in the CDW phase of the 1DSF model. The existence of a CDW order will be discussed in 
Sec.~\ref{sec:order}. 
Finally, the gap saturates at $V_{\text{BI}} \approx 22 t_{\text{s}}$ for $N_{\text{leg}}=3$ and 
$V_{\text{BI}} \approx 13 t_{\text{s}}$ for $N_{\text{leg}}=15$.  
Thus we expect to observe three possible phases in the half-filled NLM:
a one-component Luttinger liquid phase for $0 < V < V_{\text{CDW}}$, a (quasi-1D) CDW phase for $V_{\text{CDW}} < V < V_{\text{BI}}$,
and a band insulator for $V > V_{\text{BI}}$.
The investigation of the first two phases and the transition at $V_{\text{CDW}}$ will be the focus
of the next subsections.

For larger values of wire-substrate hybridization, e.g. $t_{\text{ws}}=4t_{\text{s}}$ in Fig.~\ref{fig:Ep_vs_V_Ly}(b), 
the behavior of $E_{\text{p}}$ remains qualitatively similar. The saturation coupling $V_{\text{BI}}$ increases with $t_{\text{ws}}$
and reaches $V_{\text{BI}} \approx 30 t_{\text{s}}$ for $N_{\text{leg}}=3$ and $V_{\text{BI}} \approx 17 t_{\text{s}}$ 
for $N_{\text{leg}}=15$.  
For $V<V_{\text{BI}}$ we have found that the value of $E_{\text{p}}$ decreases slightly when $t_{\text{ws}}$ increases.
This effect indicates an hybridization-induced reduction of the effective interaction in the wire, which we will discuss
in Sec.~\ref{sec:gap} for the CDW phase and in Sec.~\ref{sec:correl} for the Luttinger liquid.

Finally, we note that the overall behavior of the gap remains qualitatively similar for NLM with different numbers of legs
(as long as $N_{\text{leg}} \geq 3$ is an odd number). The value of $V_{\text{BI}}$ decreases significantly
with increasing $N_{\text{leg}}$, however.  
This decrease is due to the reduction of the effective band gap $\Delta_{\text{s}}(N_{\text{leg}})$ in the NLM,
as already found in our previous work~\cite{paper1,paper2}. 
Nevertheless, we will argue in Sec.~\ref{sec:width} that the results remain qualitatively similar
for $N_{\text{leg}} \rightarrow \infty$ and thus for the 3D wire-substrate model~(\ref{eq:hamiltonian}).

\subsection{Excitation density \label{sec:density}}

\begin{figure}[t]
\includegraphics[width=0.48\textwidth]{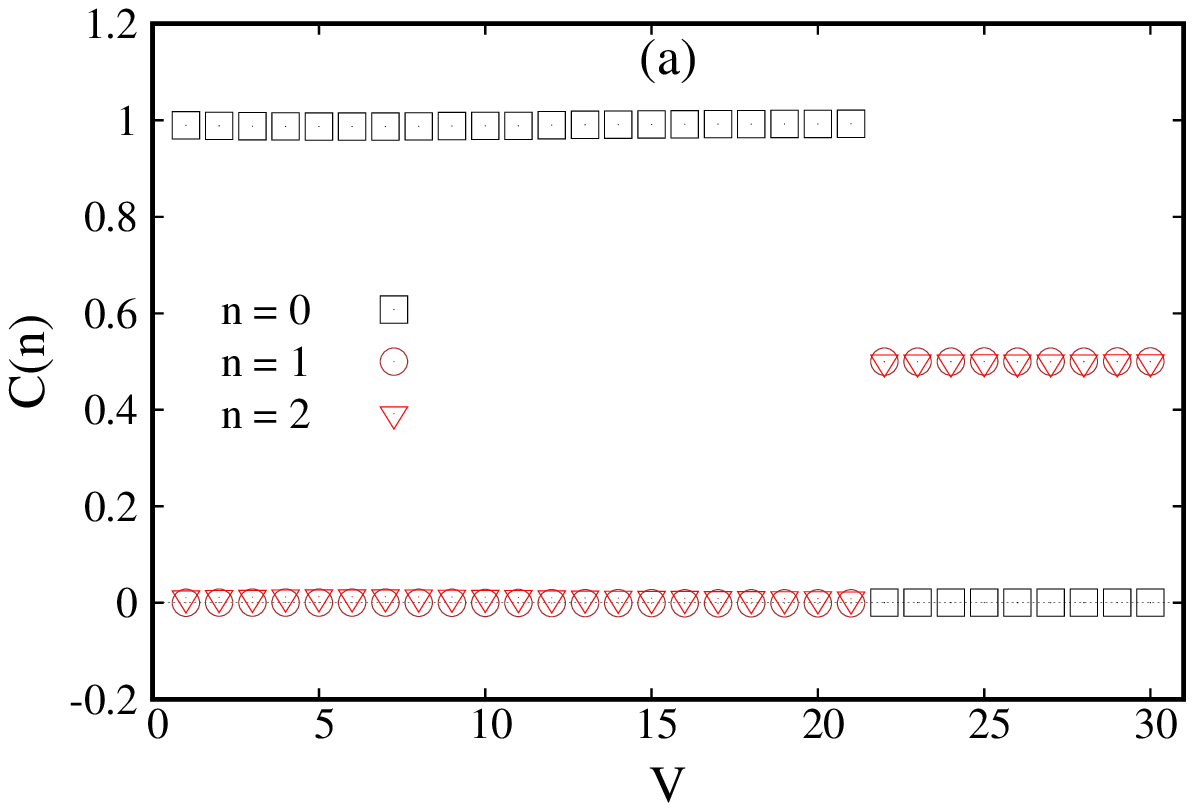}
\includegraphics[width=0.48\textwidth]{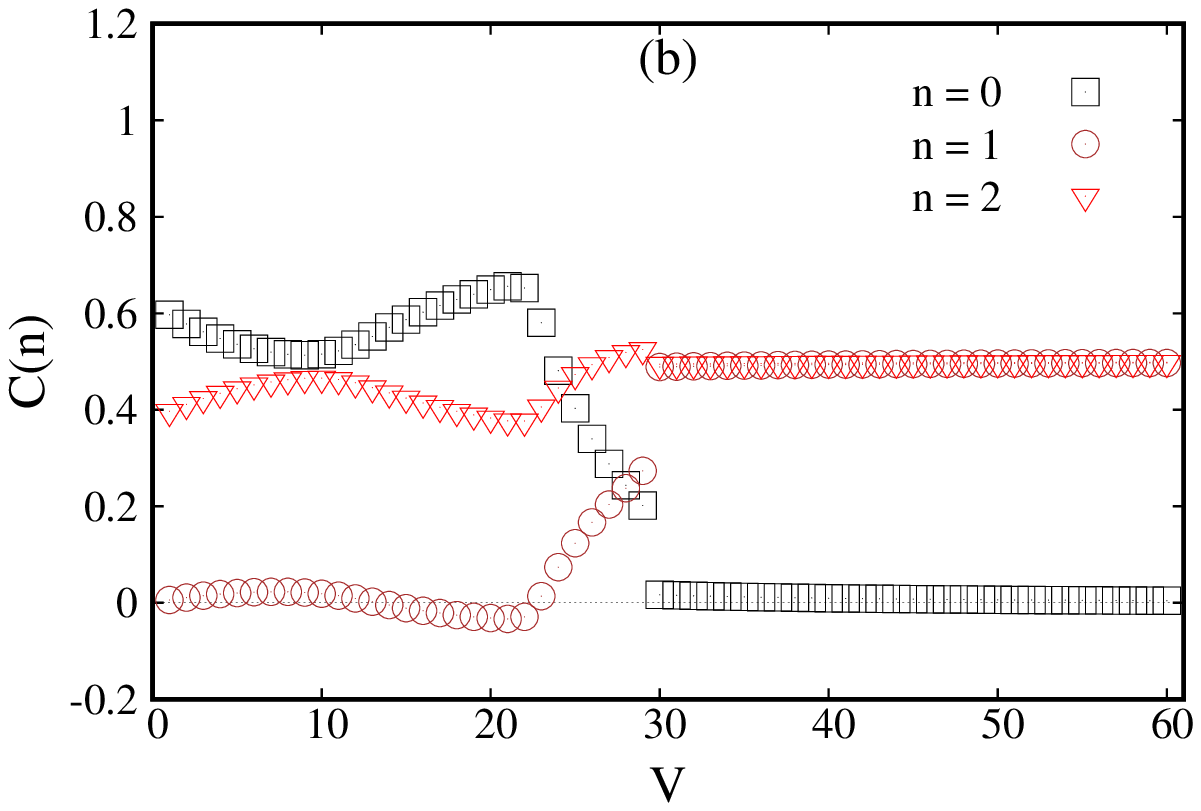}
\caption{\label{fig:chrgdens1p}
Distribution of the single-particle excitation density $C(n)$ [see Eq,~(\ref{eq:chargedens})] between the legs of a 
three-leg NLM of length $L_x=128$ with
(a) $t_{\text{ws}}=0.5$ and (b) $t_{\text{ws}}=4$.
}
\end{figure}

The existence of phases with quasi-1D low-energy excitations is confirmed by the distribution of the excess density in the legs
when one spinless fermion is added to the half-filled NLM.
This corresponds to the density distribution of the lowest single-particle excitation with the excitation energy given by Eq.~(\ref{eq:gap}).
The distribution is defined as the difference in the total density on leg $n$ between the doped and
the half-filled ground states
\begin{equation}
\label{eq:chargedens}
C(n)=\left \langle \sum^{L_x}_{x=1}g^{\dag}_{xn}g^{\phantom{\dag}}_{xn}
 \right \rangle - \frac{L_{x}}{2} \ .
\end{equation}
The expectation value is calculated for the ground-state with $N'=N+1$ spinless fermions.

The distribution is displayed in Fig.~\ref{fig:chrgdens1p} for a 3-leg NLM as a function of $V$.
For a weak hybridization $t_{\text{ws}}=0.5t_{\text{s}}$, Fig.~\ref{fig:chrgdens1p}(a) shows that
the lowest excitation (additional fermion)
is almost entirely localized on the wire for $V<V_{\text{BI}}$ but almost entirely on the substrate legs 
for $V>V_{\text{BI}}$.
For strong wire-substrate hybridization the density distribution between the legs is
more complex. 
As we can see in Fig.~\ref{fig:chrgdens1p}(b) for  $t_{\text{ws}}=4t_{\text{s}}$, 
$C(n)$ is non-monotonic as a function of $V$ and $n$. Moreover, the transition
at $V=V_{\text{BI}}$ is not so abrupt as for weak $t_{\text{ws}}$. 
Nevertheless, the lowest excitation still has a significant probability to be on the wire leg 
for $V<V_{\text{BI}}$ while it is almost entirely concentrated on the substrates legs for larger $V$. 
We have no explanation for the non-monotonic behavior.

For broader NLM with up to 15 legs, we have found that the excitation density is distributed over all legs 
for $V>V_{\text{BI}}$ as expected for the band insulator~\cite{paper2} 
but remains concentrated
on the wire leg and the first few substrate legs for weaker interactions $V$.
Therefore, the analysis of the excitation density distribution between legs confirms the localization
of the lowest excitation on or around the wire for $V<V_{\text{BI}}$ and thus the existence of an effective
1D system in the NLM as well as in the 3D wire-substrate model~(\ref{eq:hamiltonian}).

\subsection{CDW order parameter \label{sec:order}}

\begin{figure}[t]
\includegraphics[width=0.48\textwidth]{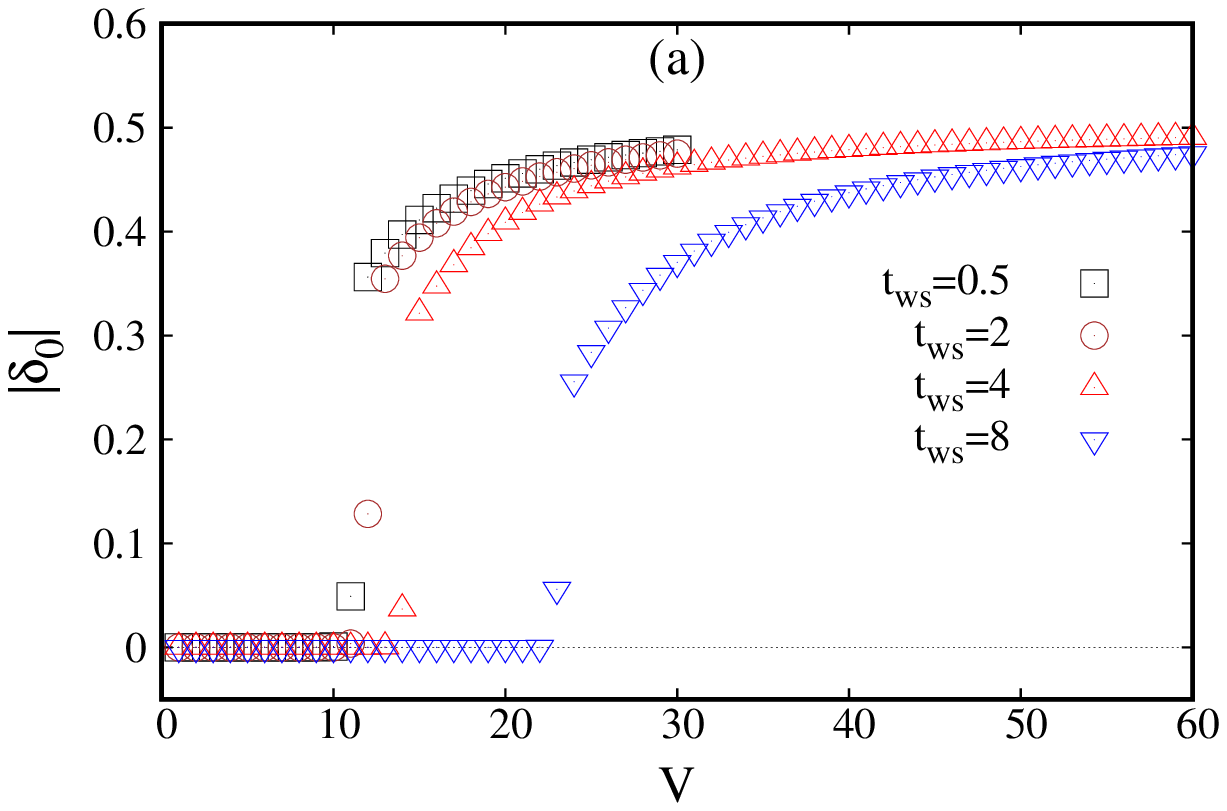}
\includegraphics[width=0.48\textwidth]{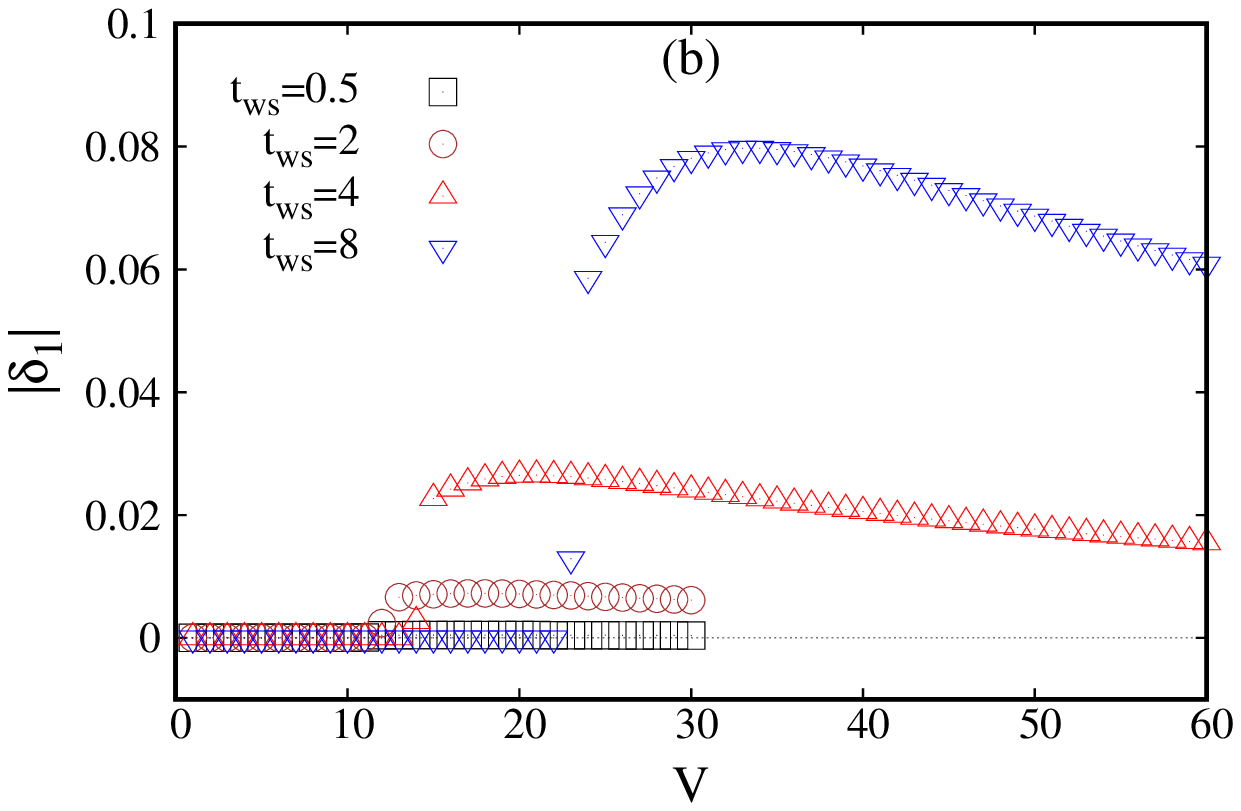}
\includegraphics[width=0.48\textwidth]{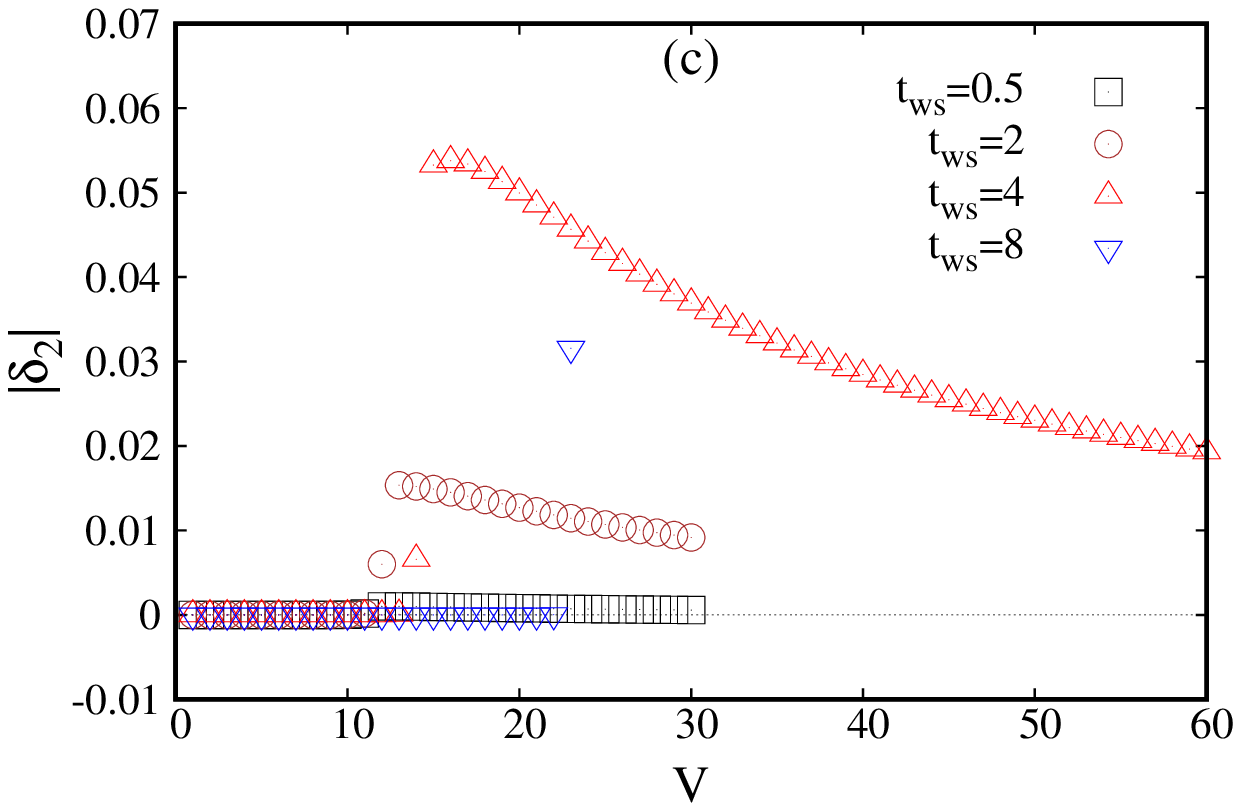}
\caption{\label{fig:chrgdensorderHF3legtws}
Absolute values of the CDW order parameter $\delta_n$ [see~Eq.~(\ref{eq:order})] on each leg ($n=0,1,2$) in a 3-leg NLM of length $L_x = 128$ as a function of 
the nearest-neighbor interaction $V$ for several values of $t_{\text{ws}}$.
}
\end{figure}

The half-filled 1DSF model exhibits long-range CDW order for $V > V_{\text{CDW}} =2t_{\text{w}}$.
In the NLM we also observe an oscillating local density for large enough coupling $V$ but the critical coupling 
$V_{\text{CDW}}$ depends on the wire-substrate hybridization.
These oscillations have the form  
\begin{equation}
  \langle g^{\dag}_{xn}g^{\phantom{\dag}}_{xn} \rangle = \frac{1}{2} + (-1)^n (-1)^x \delta_{n,x}
\end{equation}
where $\delta_{n,x}$ is a slowly varying function of $x$ due the open boundary condition in the
$x$-direction used for DMRG calculations.
Thus we can write a CDW order parameter for each leg
\begin{equation}
\label{eq:order}
  \delta_n =  \frac{1}{L_x}\sum_{x} (-1)^x \langle g^{\dag}_{xn}g^{\phantom{\dag}}_{xn} \rangle 
   = \frac{2}{L_x}\sum_{\text{even} \ x } \left ( \langle g^{\dag}_{xn}g^{\phantom{\dag}}_{xn} \rangle
   - \frac{1}{2} \right ),
\end{equation}
where the expectation value is calculated for the ground state at half filling.
Strictly, the order parameters calculated with~(\ref{eq:order}) should always vanish in a finite NLM ($L_x < \infty$).
It is a useful side-effect of the density-matrix truncation errors that DMRG reveals this broken-symmetry ground state
in finite-size ladders. However, the accuracy of $\delta_n$ is poor in the critical region, i.e., when the ladder length $L_x$
is not much longer than the correlation length.

These order parameters are shown for a 3-leg NLM and several values of $t_{\text{ws}}$ in
Fig.~\ref{fig:chrgdensorderHF3legtws}.
Clearly $\delta_n$ remains equal to zero in all legs up to a coupling that we identify as (a first estimation of) $V_{\text{CDW}}$.
Indeed, we will see in the next subsection that the single-particle gap opens close to this value as a function of $V$. 
Note that  $V_{\text{CDW}}$ increases significantly with $t_{\text{ws}}$.
Thus the wire-substrate hybridization appears to hinder the formation of a CDW ground state.
For  $V > V_{\text{CDW}}$ we observe that the order parameter on the wire $\delta_0$  increases 
monotonically with $V$. It continues to increase  in the band insulating phase ($V > V_{\text{BI}}$) and converges toward its largest possible value ($\vert \delta_0 \vert =1/2$) 
as $V \rightarrow \infty$.
This behavior shows that the wire preserves its identity as a CDW chain even in the band-insulating phase, albeit it does no longer
determine the low-energy physics of the system.

The nearest-neighbor interaction in the wire can also induce a CDW in the substrate for $V > V_{\text{CDW}}$ as shown by  
Figs.~\ref{fig:chrgdensorderHF3legtws}(b) and~(c).
As expected the sign of the order parameters $\delta_n$ alternates as $(-1)^n$.
The amplitude of the substrate CDW increases with stronger wire-substrate hybridization but is a non-monotonic function of $V$. 
It reaches a maximum at some value $V$ before decreasing gradually toward zero for $V \rightarrow \infty$. This behavior is not surprising
as a perturbation expansion in power of $t_{\text{ws}}$ yields $\delta_1 \propto t_{\text{ws}}^2/V$ for $V \gg t_{\text{ws}},  t_{\text{w}}$.
The behavior of $\delta_n$ is similar in broader NLM with up to $15$ legs.
Finally, we note that there is not any indication of the transition from the CDW phase to the band-insulating phase at $V = V_{\text{BI}}$ in the CDW order parameters. 

\subsection{Gap \label{sec:gap}}

\begin{figure}[t]
\includegraphics[width=0.47\textwidth]{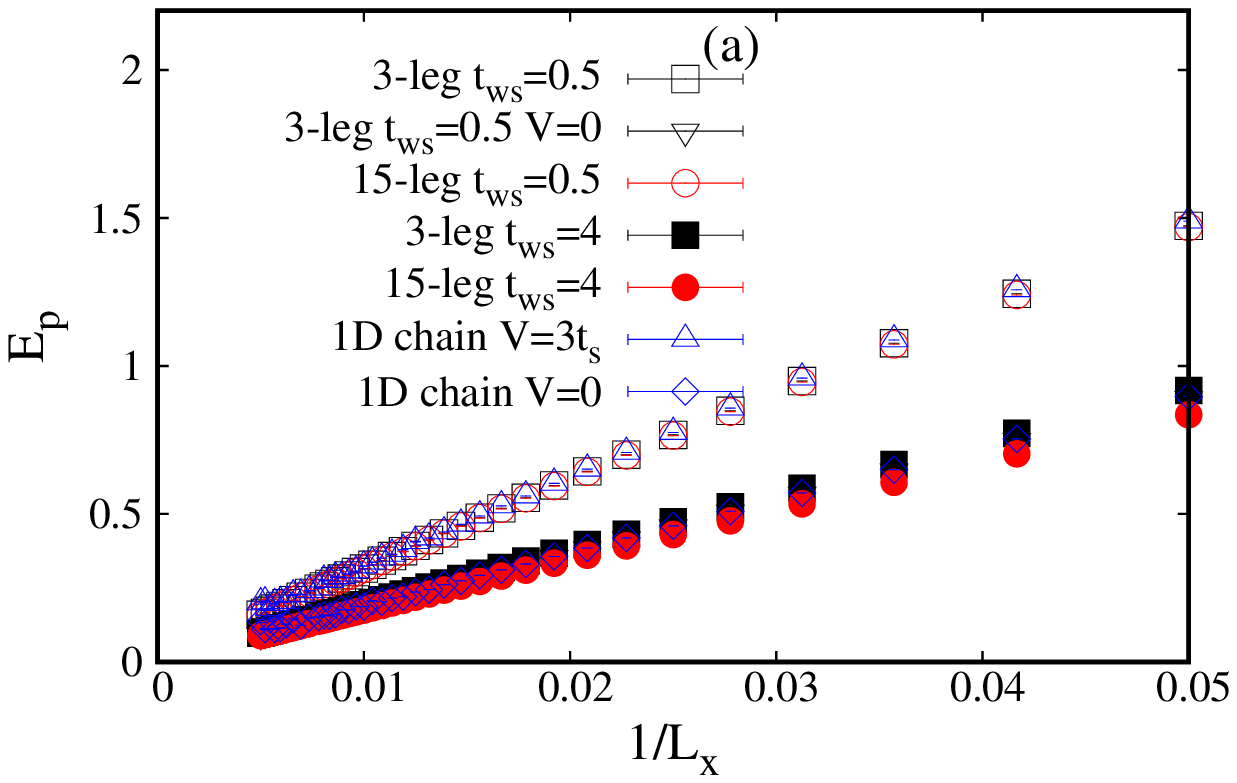}
\includegraphics[width=0.47\textwidth]{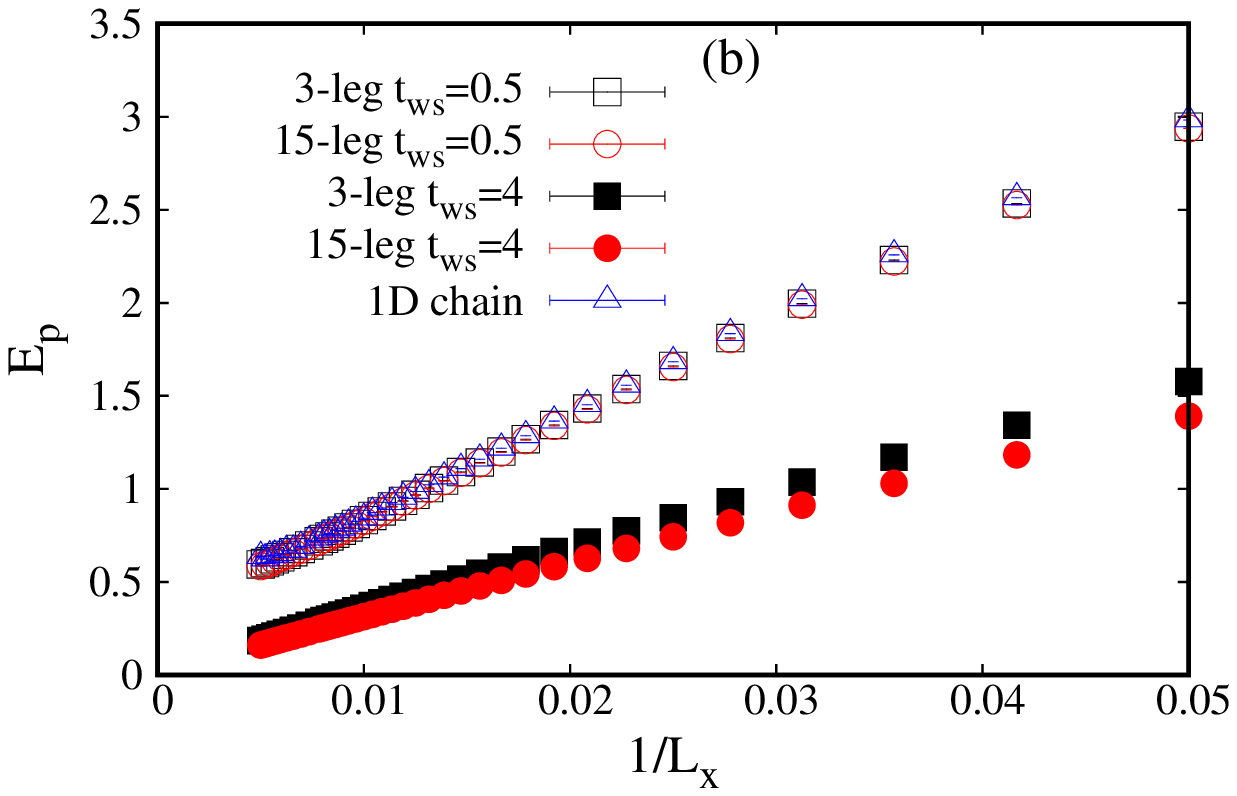}
\includegraphics[width=0.47\textwidth]{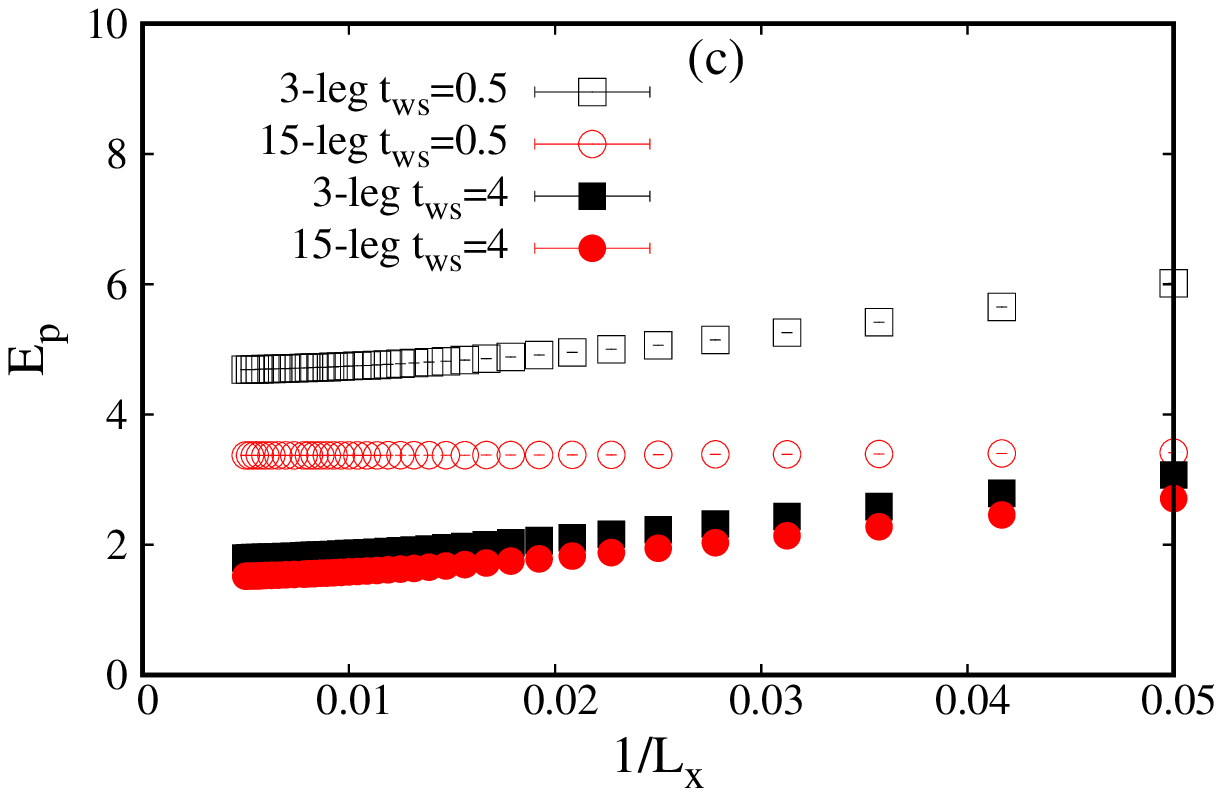}
\caption{\label{fig:EpextraptwsLy3Ly15}
Scaling of the single-particle gap $E_{\text{p}}$ as a function of the inverse ladder length $1/L_x$ for a 3-leg and a 15-leg NLM
at weak ($t_{\text{ws}}=0.5t_{\text{s}}$) and strong ($t_{\text{ws}}=4t_{\text{s}}$) wire-substrate hybridization. 
The nearest-neighbor interaction on the wire is (a) $V=3t_{\text{s}}$ (b) $V=9t_{\text{s}}$, and (c) $V=15t_{\text{s}}$.
The finite-size gaps of the 1DSF model are also shown in panels (a) and (b).
}
\end{figure}

The existence or absence of a gap in the excitation spectrum of the NLM in the thermodynamic limit can be determined by
extrapolating the finite-size gap~(\ref{eq:gap}) to the limit $1/L_x\rightarrow 0$. 
For 1D correlated conductors this extrapolation converges linearly to zero according to conformal field theory analysis~\cite{giamarchi07}.
In Fig.~\ref{fig:EpextraptwsLy3Ly15} we illustrate these extrapolations for a 3-leg and a 15-leg ladder and several values of $t_{\text{ws}}$ and $V$.

For weak interaction $V=3t_{\text{s}}$, Fig.~\ref{fig:EpextraptwsLy3Ly15}(a) shows that the extrapolations
scale linearly to zero. 
These results confirm that the lowest excitations of the NLM  are gapless in this parameter regime, which is thus 
in a Luttinger liquid phase. 
For a Luttinger liquid the slope of the linear term in $1/L_x$ in the extrapolation is proportional to the velocity $\nu^*$ of elementary excitations. 
Using $\hbar=1$ and a lattice constant $a=1$, $E_{\text{p}}(L_x) = \pi \nu/L_x$ for $L_x \gg 1$. 
We obtain $\nu\approx 9.4 t_{\text{s}}$ for $t_{\text{ws}}=0.5t_{\text{s}}$ and
$\nu\approx 6 t_{\text{s}}$ for $t_{\text{ws}}=4t_{\text{s}}$ from the data in  Fig.~\ref{fig:EpextraptwsLy3Ly15}(a).
These velocities are compatible to the exact result for the 1DSF chain~\cite{decl1966} 
\begin{equation}
\nu^* = \frac{\pi}{2} \frac{\sqrt{4t_{\text{w}}^2 - V^2}}{\arccos(V/2t_{\text{w}})}
\end{equation}
that yields $\nu^* \approx 7.8 t_{\text{s}}$ for $V=t_{\text{w}}=3t_{\text{s}}$.

We have found that the convergence of the finite-size gap toward zero and the value of the velocity
change only slightly with increasing number of legs. This confirms that these excitations
are mostly localized around the wire leg in larger NLM. Therefore,
we expect that the 3D wire-substrate model~(\ref{eq:hamiltonian}) also exhibits
these quasi-1D gapless excitations and thus that the Luttinger liquid theory
describes the low-energy physics of the wire at weak coupling $V$, even when it is hybridized with the substrate.

For intermediate interaction strength, such as $V=9t_{\text{s}}$ displayed in Fig.~\ref{fig:EpextraptwsLy3Ly15}(b),
the extrapolation leads to a finite single-particle gap in the thermodynamic limit at weak hybridization.
For  $t_{\text{ws}}=0.5t_{\text{s}}$ the extrapolated gap $E_{\text{p}} \approx 0.3 t_{\text{s}}$
is close to its value in the CDW phase of the 1DSF chain $E_{\text{p}} \approx 0.32 t_{\text{s}}$. 
The CDW order parameter $\delta_0$ is also finite
in that regime. Thus this result confirms the existence of an insulating phase with long-range CDW order
in the wire. As the scaling with $L_x$ remains almost the same for NLM with increasing number of legs, 
we expect this 1D CDW state to occur in the wire of the 3D wire-substrate model~(\ref{eq:hamiltonian}) as well.
The situation is different for strong hybridization ($t_{\text{ws}}=4t_{\text{s}}$), where one can see that
the gap vanishes linearly (within the numerical accuracy $10^{-2}t_{\text{s}}$). 
The velocity is $\nu\approx 10 t_{\text{s}}$ for the three-leg ladder and decreases slightly with increasing
number of legs down to $\nu\approx 9 t_{\text{s}}$ for $N_{\text{leg}}=15$. 
The CDW order parameter $\delta_0$ also vanishes for these parameters.
Thus for that parameter regime the NLM is still in a Luttinger liquid phase. 
This confirms that the wire-substrate hybridization increases the critical coupling $V_{\text{CDW}}$ 
that is necessary to destabilize the Luttinger liquid and to induce the long-range-ordered CDW insulator.

Figure~\ref{fig:EpextraptwsLy3Ly15}(c) shows that for the strong coupling $V=15t_{\text{s}}$ the extrapolation
results in finite values of $E_{\text{p}}$. 
For strong hybridization $t_{\text{ws}}=4t_{\text{s}}$ 
the extrapolated gap diminishes only slightly with increasing number of legs (from $E_{\text{p}} \approx 1.7 t_{\text{s}}$ 
for the 3-leg ladder to $E_{\text{p}} \approx 1.4 t_{\text{s}}$ for the 15-leg ladder) and   
the CDW order parameter $\delta_0$ is finite.
Thus this case corresponds again to a CDW phase of the NLM. However, the CDW gap 
in the NLM is now clearly smaller 
than its value in the 1DSF chain $E_{\text{p}} \approx 4.7t_{\text{s}}$
This observation and the shifting of the critical coupling $V_{\text{CDW}}$ to higher values
show that the effective CDW-inducing interaction in the wire is reduced by the wire-substrate hybridization
compared to the bare value $V$.
For weak hybridization ($t_{\text{ws}}=0.5t_{\text{s}}$), however, the extrapolated gap diminishes
significantly with increasing $N_{\text{leg}}$ until it reaches
the effective band gap $\Delta_{\text{s}}(N_{\text{leg}})$.
This case clearly corresponds to the band-insulating phase as confirmed by Fig.~\ref{fig:Ep_vs_V_Ly}(a).

\subsection{Correlation functions \label{sec:correl}}

\begin{figure}[t]
\includegraphics[width=0.48\textwidth]{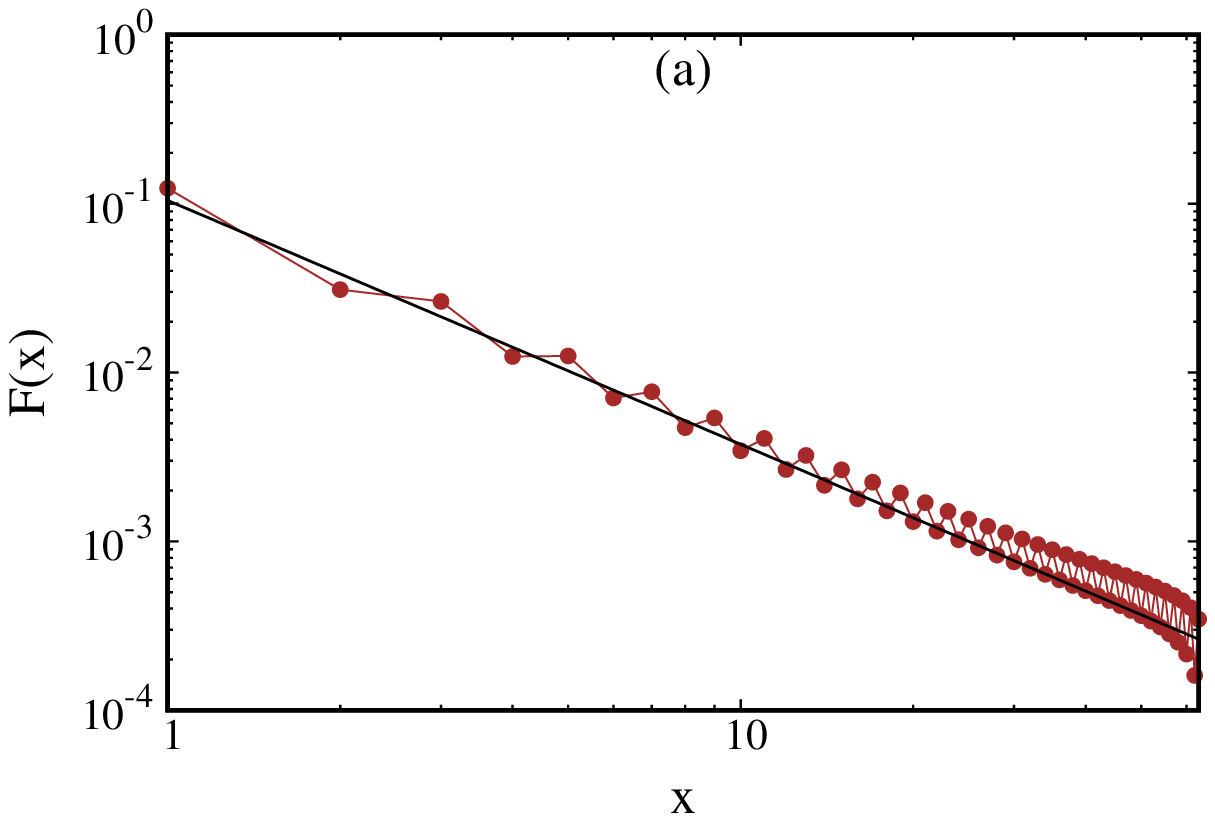}
\includegraphics[width=0.48\textwidth]{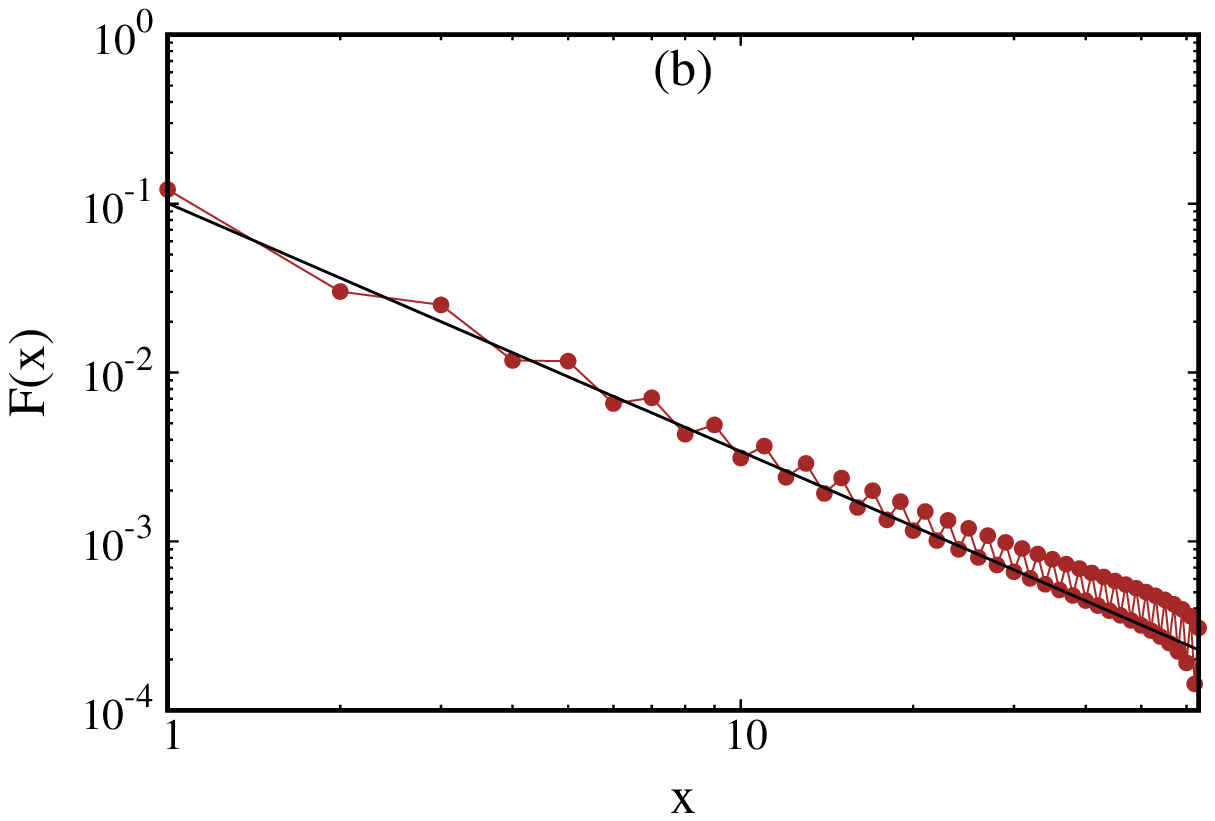}
\caption{\label{fig:Correlations}
Density-density correlation functions $F(x)$ on the wire as a function of distance $x$ in the leg direction
for (a) a three-leg NLM and (b) a 15-leg NLM with $t_{\text{ws}}=2t_{\text{s}}$ and $V=4t_{\text{s}}$.
Straight lines indicate the power-law fit $\vert F(x) \vert \propto \vert x \vert^{-2K}$.
}
\end{figure}

In the previous subsections, we have shown that the Luttinger liquid and CDW phases are robust against the wire-substrate hybridization
in the NLM. The effective interaction leading to the CDW state seems to be weaker than the bare interaction $V$, however.
In the Luttinger liquid phase the strength of correlations should be reflected in the behavior of correlation functions~\cite{Schoenhammer,giamarchi07}.  
A Luttinger liquid is characterized by a power-law decay of the density-density correlations with an exponent
that depends on the interaction strength.
The density correlations on the wire is defined as
\begin{eqnarray}\label{eq:correlation}
  F(x_0-x) &=& \left \langle g^{\dag}_{x_0\text{w}}g^{\phantom{\dag}}_{x_0\text{w}}
  g^{\dag}_{x\text{w}}g^{\phantom{\dag}}_{x\text{w}} \right \rangle \\ \nonumber
 &&- \left \langle g^{\dag}_{x_0\text{w}}g^{\phantom{\dag}}_{x_0\text{w}} \right \rangle
 \left \langle g^{\dag}_{x\text{w}}g^{\phantom{\dag}}_{x\text{w}}  \right \rangle 
\end{eqnarray}
where $x_0=\frac{L_x}{2}$ is chosen to be in the middle of the ladder for our DMRG computations
and the expectation values are calculated for the ground-state of the half-filled NLM.
According to the Luttinger liquid theory these density correlations decay asymptotically as 
\begin{equation}\label{eq:exponent}
 F(x) \sim \frac{\cos(2k_{\text{F}}x)}{x^{2K}}
\end{equation}
where $K$ is the Luttinger liquid parameter.
In Fig.~\ref{fig:Correlations} we display the absolute values of the density-density correlation functions measured
on the wire for a 3-leg and a 15-leg NLM with $t_{\text{ws}}=2t_{\text{s}}$ and $V=4t_{\text{s}}$. 
The overall power-law decay is clearly visible although the ladder length is finite.

\begin{figure}[t]
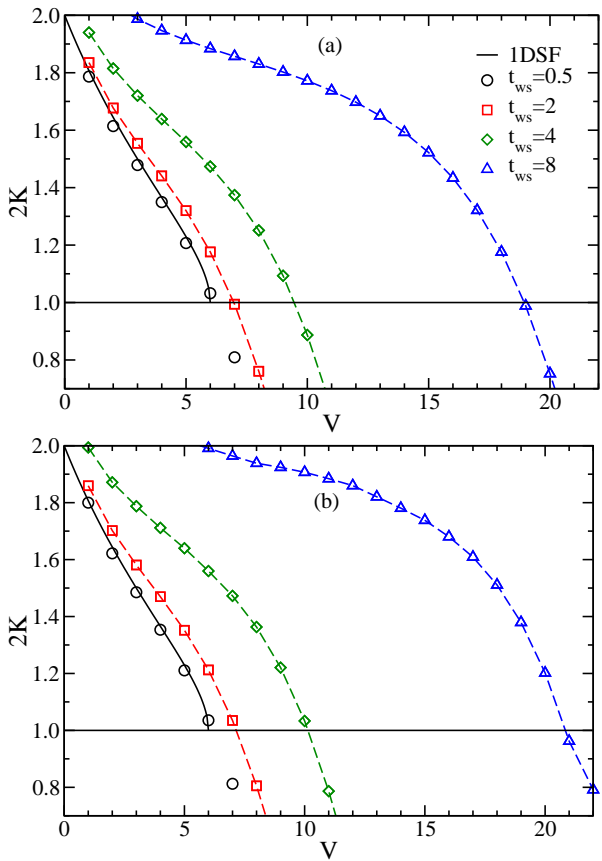

\includegraphics[width=0.44\textwidth]{fig6a}
\includegraphics[width=0.44\textwidth]{fig6b}
\caption{\label{fig:V_exponents}
Exponent $2K$ calculated with DMRG as a function of the bare nearest-neighbor interaction $V$ for several values
of the wire-substrate hybridization $t_{\text{ws}}$ in (a) a three-leg NLM and (b) a 15-leg NLM.
Solid lines show the exact result for the 1DSF model~(\ref{eq:K_vs_V}).
Dashed lines are guides for the eye.
}
\end{figure}

Thus the exponent $2K$ in Eq.~(\ref{eq:exponent}) can be extracted from the correlation functions
calculated with DMRG for finite length $L_x$ using a simple power-law fit $\vert F(x) \vert \propto \vert x \vert^{-2K}$ . 
The results are plotted against the bare interaction $V$ in Fig.~\ref{fig:V_exponents} 
for a 3-leg and a 15-leg with several values of $t_{\text{ws}}$. 
From the Bethe Ansatz solution of the 1DSF model we know that the Luttinger liquid parameter $K$ is given by~\cite{giamarchi07} 
\begin{equation}\label{eq:K_vs_V}
K = \frac{\pi}{2}\frac{1}{\pi - \arccos(V/2t_{\text{w}})}
\end{equation}
in the limit $t_{\text{ws}}=0$.
This relation is approximately fulfilled for weak wire-substrate hybridizations but obviously not for strong ones.

Figure~\ref{fig:V_exponents} shows clearly that the exponent decreases with increasing $V$
from the exact result $K=1$ for an noninteracting chain ($V=0$)
to the value $K=\frac{1}{2}$. Field theory predicts that a transition to an insulating CDW state occurs at this value of $K$ in a half-filled 1DSF system~\cite{giamarchi07}.
Therefore, we can determine the critical coupling $V_{\text{CDW}}$ from our DMRG calculations using the condition $K=\frac{1}{2}$ for
the fitted correlation function exponents. 

This procedure yields values for $V_{\text{CDW}}$ that are 
smaller than the ones estimated from the vanishing of the order parameter $\delta_0$ given by Eq.~(\ref{eq:order}).
We note that the extrapolated gaps in Sec.~\ref{sec:gap} as well as the central charges determined from the entropy 
in the next subsection agree better with the values of $V_{\text{CDW}}$ calculated from the Luttinger exponents
than with those determined from the order parameter. This is not surprising as the calculation of $\delta_0$ is quite inaccurate in the critical region
as discussed in~Sec.~\ref{sec:order}.

It is also clear in Fig.~\ref{fig:V_exponents} that $K$ and $V_{\text{CDW}}$ increase with $t_{\text{ws}}$.
Thus the wire-substrate hybridization reduces the effective interaction in the wire in comparison to the bare interaction $V$
of the isolated wire and stabilizes the Luttinger liquid phase with respect to the formation of the insulating CDW state.
Finally, we note again that the results vary only slightly with the NLM width $N_{\text{leg}}$.

Therefore, both the gap extrapolations to the thermodynamic limit and the power-law decay of correlation functions
demonstrate the existence of a Luttinger liquid phase in the NLM.
In principle, one could check further properties of Luttinger liquids such as the power-law behavior of the local density of states~\cite{jeck13}
or the renormalization of the linear conductance~\cite{bisc17} using DMRG. The density of states has the advantage of being experimentally accessible 
using scanning tunneling spectroscopy. The computational cost of such computations is much higher than for the quantities calculated in the present work, however.
Thus these properties will be investigated in further works.

\subsection{Scaling of entropy\label{sec:entropy}}

\begin{figure}[t]
\includegraphics[width=0.48\textwidth]{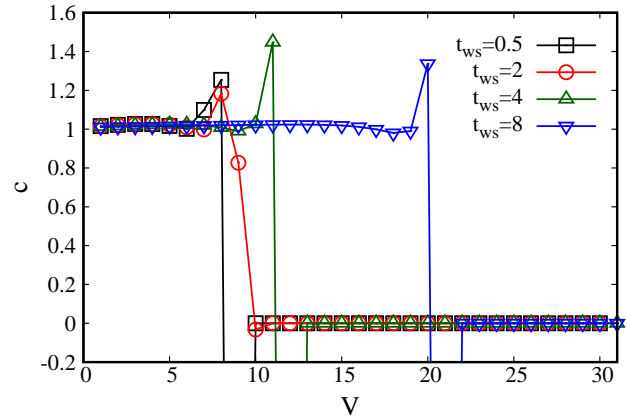}
\caption{\label{fig:central_charge_vs_tperp}
Central charge of a three-leg NLM calculated with Eq.~(\ref{eq:charge}) as a function of $V$ for several $t_{\text{ws}}$.
}
\end{figure}

The von Neumann entanglement entropy $S$ is defined using the reduced density matrix $\rho_{A}$
of a subsystem $A$ 
\begin{equation}
 S = -\text{Tr}\rho_{A}\ln\rho_{A}.
\end{equation}
For critical 1D systems with open-boundary conditions and length $L_x$ that are divided 
 into two pieces of length $l_x$ and $L_x-l_x$,
 the entropy fulfills the relation
\begin{equation}
\label{eq:entropy}
 S(L_x,l_x) = \frac{c}{6} \ln\left[ \frac{2L_x}{\pi}\sin\left(\frac{\pi l_x}{L_x}\right) \right] + c_1 + s_b,
\end{equation}
where $c$ is the central charge, $c_1$ is a non-universal constant, and $s_b$ is the boundary entropy~\cite{entropy}.
The central  charge is a useful information extracted from the finite-size scaling of the entropy
because it exhibits a discontinuous behavior at the critical point between
the Luttinger liquid phase and the CDW phase in the 1DSF model, where it jumps from one to zero.

We used the entanglement entropy obtained with DMRG for NLM with different lengths $L_x$ and $L_x^{\prime}$ to calculate 
\begin{equation}
\label{eq:charge}
 c(L_x,L_x^{\prime}) = 6 \left[ \frac{S(L_x,L_x/2)-S(L_x^{\prime},L_x^{\prime}/2)}{\ln(L_x)-\ln(L_x^{\prime})}\right] .
\end{equation}
This quantity should approximate the central charge if Eq.~(\ref{eq:entropy}) is valid for a three-leg NLM with (at most) one gapless excitation 
mode. Note that this is not the best approach to compute the central charge with DMRG  because of complicated finite-size and boundary 
effects~\cite{nish11} but it is sufficient for our purpose. Yet we confine this calculation to three-leg NLM.

Figure~\ref{fig:central_charge_vs_tperp} shows our results as a function of
$V$ using three-leg
NLM of lengths $L_x=500$ and $L^{\prime}_x=300$.
The central charge appears to jump abruptly from a finite value close to one at weak couplings $V$ to zero for larger couplings $V$.
The positions of these jumps agree with the critical couplings $V_{\text{CDW}}$ determined from the 
power-law exponents for density-density correlation functions in the previous subsection.
Close to the critical couplings $V_{\text{CDW}}$ our results are less accurate because the correlation lengths can exceed the system sizes
used (up to $L_x=500$). Eq.~(\ref{eq:charge}) may even return negative values because the DMRG approximation to the ground state in the shorter ladder 
is in the gapless phase and thus is more entangled than the ground state of the longer ladder, which is the CDW phase.

\subsection{Influence of the ladder width \label{sec:width}}

As explained in Sec~\ref{sec:nlm} the NLM properties should agree with those of the 3D wire-substrate model~(\ref{eq:hamiltonian})
when the odd number of legs $N_{\text{leg}}$ is large enough.
Therefore, it is important to understand how these properties changed with increasing $N_{\text{leg}}$.
In the present work we have found that ground-state properties measured close to the wire and low-energy excitations 
vary very little with $N_{\text{leg}}$ when $V < V_{\text{BI}}$, i.e. in the Luttinger liquid and CDW phases.
They vary significantly in the band insulator phase for $V > V_{\text{BI}}$, however. For instance, the gap $E_{\text{p}}$
[$= \Delta_{\text{s}}(N_{\text{leg}})$ in this phase] decreases as shown in Fig.~\ref{fig:Ep_vs_V_Ly}.

\begin{figure}[t]
\includegraphics[width=0.48\textwidth]{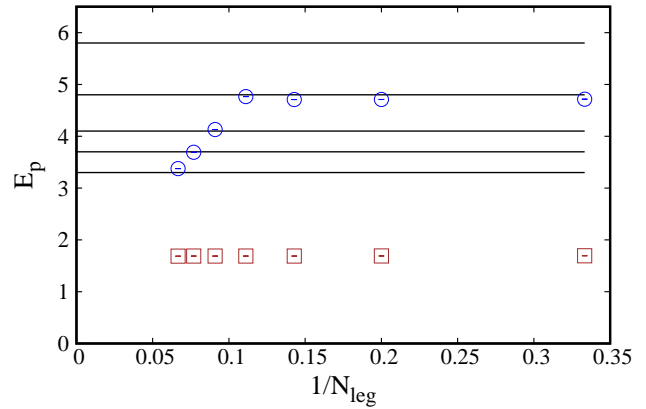}
\caption{\label{fig:gap_nleg}
Single-particle gap $E_{\text{p}}$ as a function of the number of 
legs $N_{\text{leg}}$ in the NLM for $V=11t_{\text{s}}$ (squares) and $15t_{\text{s}}$ (circles) both with $t_{\text{ws}}= 0.5t_{\text{s}}$.
Horizontal lines indicate the effective band gaps $\Delta_{\text{s}}(N_{\text{leg}})$ for noninteracting NLM with 
$N_{\text{leg}}=7,9,11,13$ and $15$ from top to bottom.
}
\end{figure}

The main problem with the finite NLM width is that the phase boundary $V_{\text{BI}}$ moves to significantly
lower values when $N_{\text{leg}}$ increases because it is determined by the effective band gap $\Delta_{\text{s}}(N_{\text{leg}})$.
Thus a NLM with a fixed coupling $V$ can jump from the CDW insulator phase to the band insulator phase at some finite $N_{\text{leg}}$.
This is illustrated in Fig.~\ref{fig:gap_nleg} where we show the single-particle gap $E_{\text{p}}$
as a function of the number of legs for two values of $V$. The three-leg NLM is in the CDW phase for both couplings $V$
as $E_{\text{p}} < \Delta_{\text{s}}(N_{\text{leg}}=3) \approx 10.4 t_{\text{s}}$.
For $N_{\text{leg}} \gg 1$ only the NLM with the weaker coupling ($V= 11t_{\text{s}}$) remains in the CDW phase
because its gap $E_{\text{p}}$ remains smaller than 
$\lim_{N_{\text{leg}}\rightarrow \infty} \Delta_{\text{s}}(N_{\text{leg}}) = \Delta_{\text{s}} = 2t_{\text{s}}$ 
while the NLM with the stronger coupling ($V= 15t_{\text{s}}$) moves to the band insulator phase for $N_{\text{leg}} > 9$.

Therefore, it is difficult to determine the true extent of the quasi-1D CDW phase in the 3D model~(\ref{eq:hamiltonian})
from the NLM results. For the Luttinger liquid phase, however, the phase boundary ($V_{\text{CDW}}$) varies little 
with $N_{\text{leg}}$ and thus we conclude that there is an extended parameter regime for which the low-energy physics
of the 3D model~(\ref{eq:hamiltonian}) is determined by a Luttinger liquid localized around the wire.

\section{Conclusion}

We have investigated an interacting one-dimensional spinless fermion system coupled to a three-dimensional insulating substrate using a mapping to
narrow ladder models (NLM) and the DMRG method. We were able to investigate the 
NLM properties as a function of the ladder width systematically. Our results confirm that the NLM approach is
a useful tool to investigate correlated atomic wires deposited on semiconducting substrates.

We have found that the low-energy excitations are localized on or close to the wire for a wide range of model parameters
and that the system exhibits the low-energy properties of Luttinger liquids or quasi-one-dimensional CDW insulators in this regime.
These one-dimensional phases occur even for strong hybridization between wire and substrate. 
Thus our study confirms that 
Luttinger liquids and one-dimensional CDW insulators are not always destroyed by the coupling to their three-dimensional environment.
Therefore, these phases could occur in the low-energy properties of atomic wires deposited on semiconducting substrates.

\begin{acknowledgments}
A. Abdelwahab thanks M. Rizzi for useful discussions.
This work was done as part of the Research Units \textit{Metallic nanowires on the atomic scale: Electronic
and vibrational coupling in real world systems} (FOR1700) of the German Research Foundation (DFG) and was supported by
grant~JE~261/1-2. 
The DMRG calculations were carried out on the cluster system
at the Leibniz University of Hannover.
\end{acknowledgments}

  \end{document}